# Properties of Nb$_3$Sn films fabricated by magnetron sputtering from a single target

Md. Nizam Sayeed[1,2], Uttar Pudasaini[3], Charles E. Reece[3], Grigory V. Eremeev[4], and Hani E. Elsayed-Ali[1,2*]

[1]*Department of Electrical and Computer Engineering, Old Dominion University, Norfolk, VA 23529, USA*
[2]*Applied Research Center, 12050 Jefferson Avenue, Newport News, VA 23606, USA*
[3]*Thomas Jefferson National Accelerator Facility, Newport News, VA 23606, USA*
[4]*Fermi National Accelerator Laboratory, Batavia, IL 60510, USA*

*Abstract*

Superconducting Nb$_3$Sn films were fabricated on sapphire and fine grain Nb substrates by magnetron sputtering from a single stoichiometric Nb$_3$Sn target. The structural, morphological and superconducting properties of the films annealed for 24 h at temperatures of 800-1000 °C were investigated. The effect of the annealing time at 1000 °C was examined for 1, 12, and 24 h. The film properties were characterized by X-ray diffraction, scanning electron microscopy, atomic force microscopy, energy dispersive X-ray spectroscopy, and Raman spectroscopy. The DC superconducting properties of the films were characterized by a four-point probe measurement down to cryogenic temperatures. The RF surface resistance of films was measured over a temperature range of 6-23 K using a 7.4 GHz sapphire-loaded Nb cavity. As-deposited Nb$_3$Sn films on sapphire had a superconducting critical temperature of 17.21 K, which improved to 17.83 K when the film was annealed at 800 °C for 24 h. For the films annealed at 1000 °C, the surface Sn content was reduced to ~11.3 % for an annealing time of 12 h and to ~4.1 % for an annealing time of 24 h. The Raman spectra of the films confirmed the microstructural evolution after annealing. The RF superconducting critical temperature of the as-deposited Nb$_3$Sn films on Nb was 16.02 K, which increased to 17.44 K when the film was annealed at 800 °C for 24 h.

## 1. Introduction

Nb$_3$Sn is of interest as a coating for superconducting radiofrequency (SRF) cavities due to its higher critical temperature $T_c$ of ~18.3 K and superheating field $H_{sh}$ of ~400 mT [1]. Nb$_3$Sn cavities have the potential to achieve high quality factor $Q_0$ when operated at 4 K and can replace the bulk Nb cavities that are operated at 2 K [2,3]. The higher operating temperature of Nb$_3$Sn cavities compared to Nb can significantly reduce the operating cost. However, Nb$_3$Sn cannot be used as bulk material due to its fragile nature and poor thermal conductivity. Nb$_3$Sn thin films coated on niobium or copper are considered as potential alternative materials for SRF cavities [2,3]. A 1.3 GHz single-cell Nb$_3$Sn/Nb cavity fabricated by Sn vapor diffusion at Jefferson Lab have demonstrated a $Q_0 \geq 2 \times 10^{10}$ at 4 K before quenching at a field $\geq$ 15 MV/m [4], while Nb$_3$Sn/Nb CEBAF five-cell cavities had a low-field $Q_0$ of ~$3 \times 10^{10}$ [5] and maximum accelerating gradient up to 15 MV/m at 4 K [6]. An accelerating gradient of 22.5 MV/m at 4 K was achieved at Fermilab for a 1.3 GHz single-cell cavity [7]. At Cornell University, Nb$_3$Sn-coated 2.6 GHz cavities had a quality factor of $8 \times 10^9$ at 4 K, which is 50 times higher than that of 2.6 GHz Nb cavities at 4 K, and preliminary data on Nb$_3$Sn-coated 3.9 GHz cavities showed a low field $Q_0$ of ~$2 \times 10^9$ at 4.2 K [8]. Nb$_3$Sn films on sapphire substrates also have potential applications for superconducting microwave devices [9].

Magnetron sputtering has been used to grow Nb$_3$Sn films from a single stoichiometric Nb$_3$Sn target [10-13], using separate Nb and Sn targets to deposit multilayers that are thermally interdiffused [14-16], or co-sputtering of Nb and Sn [17]. Early successful fabrication of Nb$_3$Sn films by magnetron sputtering from a stoichiometric target was performed at Argonne National Lab [10, 11]. There, Nb$_3$Sn films were grown on sapphire and the superconducting properties were examined for films grown at various sputter argon background pressures of 5 to 50 mTorr and substrate temperatures of 600 to 800 °C. The reported $T_c$ and critical current density $J_c$ of the grown films were up to 18.3 K and $15 \times 10^6$ A/cm$^2$, respectively [11]. Nb$_3$Sn films sputtered from a single phase Nb$_3$Sn target on MgO substrates had a $T_c$ of 15.3 K after annealing at 900 °C for 1 h [12]. The formation of Nb$_3$Sn by the reactive interdiffusion of Nb/Sn multilayers on oxidized Si and sapphire at high temperatures was investigated at AT&T Bell Laboratories [14]. The experiments of Nb/Sn multilayers on oxidized Si and sapphire showed that, Nb$_6$Sn$_5$ phase was present during the growth of A15 Nb$_3$Sn. The Nb$_6$Sn$_5$ phase disappeared above 800 °C. In this study, the highest $T_c$ of 17.45 K was observed for films annealed at 850 °C. Nb$_3$Sn sputtered on copper [13,18,19] and Nb [20, 21] substrates were studied for its application in SRF cavities. For Nb$_3$Sn deposited on copper, the annealing temperature was limited to 830 °C due to a lower melting point of copper and the mismatch between the thermal expansion coefficients of copper and Nb$_3$Sn [13]. Niobium has a high melting point and a thermal expansion coefficient close to that of Nb$_3$Sn [22]. Therefore, the deposition of Nb$_3$Sn on Nb and annealing of the grown film can be performed at temperatures above 830 °C. Nb$_3$Sn films grown by magnetron sputtering on different substrates have shown DC superconducting $T_c$ of up to 15.7 K for copper [18], 17.7 K for niobium [21], and 17.93 K for sapphire substrates [23]. For SRF cavity applications, the RF properties of Nb$_3$Sn films are of interest. The RF properties of magnetron sputtered Nb$_3$Sn films on sapphire [24, 25] and copper substrates [26] were reported previously.

We present the impact of annealing temperatures over the temperature range of 800-1000 °C on the material, DC, and RF superconducting properties of Nb$_3$Sn films grown on sapphire and Nb by magnetron sputtering from a single stoichiometric Nb$_3$Sn target. The changes in film structure, morphology, and chemical composition due to different annealing conditions on different substrates is compared with the DC and RF superconducting properties of the films. We established the growth conditions to produce Nb$_3$Sn films with high $T_c$ and very sharp superconducting transitions on sapphire substrates and applied the conditions to grow Nb$_3$Sn films on niobium substrates. The RF surface resistance of the Nb$_3$Sn films sputtered on Nb was then characterized with a 7.4 GHz sapphire-loaded Nb cavity.

## 2. Experimental details

Nb$_3$Sn was deposited on sapphire and Nb substrates by DC magnetron sputtering from a stoichiometric Nb$_3$Sn target made by co-evaporation from 2 sources (Kurt J. Lesker EJTNBSN302A4, 99.9% purity). The substrates were 430 μm thick double-side polished sapphire with C-M orientation (University Wafers Inc. part # 1251). The substrates were cleaned with ethanol and isopropanol and dried with N$_2$ gas flow before loading in the deposition chamber. The 1 cm × 1 cm Nb substrates were cut from a Nb slice (Tokyo Denkai Co., Japan, RRR ~250) by wire electrical discharge machining and then buffered chemical polished (BCP in 1:1:1 volume ratio of 49% HF, 70% HNO$_3$ and 85% H$_3$PO$_4$) to remove 100 μm. The Nb substrates were baked at 800 °C for 2 h in a vacuum furnace operating at or below 10$^{-5}$ Torr. A second BCP was applied to etch an additional 25 μm from the surface.

The sputtering system used is ATC Orion-5 sputter coater from AJA International. The sample chamber was evacuated by a turbomolecular pump to low 10$^{-7}$ Torr before deposition. The deposition was done at 3 mTorr Ar background pressure at a flow rate of 20 SCCM at a constant magnetron current of 150 mA. The substrate holder was rotated at 30 rpm during the deposition to ensure uniform coating. During the deposition, the substrates were heated at ~800 °C using a quartz heater. Samples sputtered on sapphire were deposited for 6 h, while samples sputtered on Nb were deposited for 14 h, all under the same magnetron operating current, the same Ar background pressure, and fixed target-to-substrate distance of 10 cm. The coated samples were removed from the sputter coater and then annealed for 24 h at 800, 900, and 1000 °C in a separate vacuum furnace, described in [6]. For films annealed at 1000 °C, some samples were also annealed for 1 and 12 h to study the effect of annealing time on film properties. The pressure during the annealing was ~10$^{-6}$ Torr and a temperature ramp rate of 5 °C/min was used.

X-ray diffraction (XRD) patterns of the films were obtained from Rigaku Miniflex II X-ray diffractometer with Cu-Kα radiation. Raman spectra of the films were obtained using a Renishaw inVia Raman microscope and spectrometer. The excitation source was an Ar ion laser at 514 nm wavelength. The laser spot has 2.5 μm diameter at its focal point. The Raman spectra were recorded at three different spots of each sample. The optical images of the film surface were captured by a HIROX RH-2000 high resolution digital video microscope. The film morphology was examined by field-emission scanning electron microscopic images (FESEM S-4700, Hitachi, Japan). A Noran 6 energy dispersive X-ray spectroscopy (EDX) detector connected to a Jeol JSM 6060 LV scanning electron microscope was used to measure the surface elemental composition. EDX was performed using 15 kV accelerating voltage on a surface covering ~1.2 mm$^2$ area at five randomly-selected regions of each sample. The X-ray transmission fraction for Nb-L$\alpha$, and Sn-L$\alpha$ lines estimated from the Anderson-Hasler equation is 0.68 µm and 0.76 µm, respectively, for 15 keV electrons [24]. The surface roughness of the samples was measured by a Digital Instrument Dimension 3100 Atomic Force Microscope (AFM) operating in tapping mode. Film root-mean-square (RMS) roughness was obtained over 5 × 5 µm$^2$ scan size at 3 different areas of the films. The $T_c$ was measured by four-point probe resistance measurement down to a temperature below $T_c$ of the studied film using an isothermal multi-sample measurement system with drive current of 1 mA. The four-point probe measurement system uses calibrated Cernox thermometers with a sensitivity of 50 mK to sense the temperature of the surface of the film [27]. The Cernox thermometers were placed on a copper sample board where the samples were mounted. $T_c$ was calculated from the mean value of $T^{90}$ and $T^{10}$ (temperature when the resistance is 90% and 10% of the final resistance before the transition, respectively). The superconducting transition width $\Delta T_c$ was calculated from the difference between $T^{90}$ and $T^{10}$. The residual resistance ratio ($RRR$) was measured from the ratio of resistance at 300 K and 20 K. RF surface resistance of the Nb$_3$Sn films coated on 2-inch diameter Nb disks were measured inside a 7.4 GHz sapphire-loaded Nb cavity described in [28]. The sample-loaded cavity was placed inside a dewar providing liquid He at a pressure of 23 Torr to maintain a temperature of 2 K at the cavity surface. The sample placed inside the cavity was thermally isolated from the cavity and connected to a calorimeter. Therefore, the temperature on the sample can be varied by conducting heat via the calorimeter, whereas the cavity remains at 2 K. The sample temperature was varied from 6 to 22 K, which is a sufficient range to measure the transition due to both Nb ($T_c$ = 9.3 K) and Nb$_3$Sn ($T_c$ = 18 K). Cernox thermometers with a sensitivity of 50 mK were connected at the back of the substrates to measure the temperature.

## 3. Results and Discussion
### 3.1. Nb$_3$Sn deposited on sapphire
#### 3.1.1. Film structure, morphology, and composition

Fig. 1 shows the XRD patterns of the as-deposited film and films annealed at different temperatures for 24 h. As-deposited films showed both Nb and Nb$_3$Sn diffraction peaks. The Nb peaks disappeared when the film is annealed at 800 and 900 °C. However, a few Nb peaks reappeared when the film was annealed at 1000 °C for 24 h. For films annealed at 800 and 900 °C, the Nb$_3$Sn peak intensity became stronger, indicating better crystallinity. At an annealing temperature of 1000 °C, potential Sn evaporation resulted in weak Nb$_3$Sn diffraction peaks accompanied by intense Nb peaks indicated the degradation of the Nb$_3$Sn. An expanded view of the diffraction patterns taken for $2\theta$ of 45-75° is shown in the inset of Fig. 2. The Nb$_3$Sn (222), (320), (321), and (400) diffraction peaks are observed. The peak intensity increased with the annealing temperature for annealing from 800 to 900 °C, then decreased for annealing at 1000 °C. The XRD pattern of the film annealed at 1000 °C for 24 showed diffraction peaks corresponding to Nb (110), (200), (211), and (220). All annealed films showed two diffraction peaks near $2\theta$ of 37° and 78° that are either due to the presence of the $Nb_6Sn_5$ or $NbSn_2$.

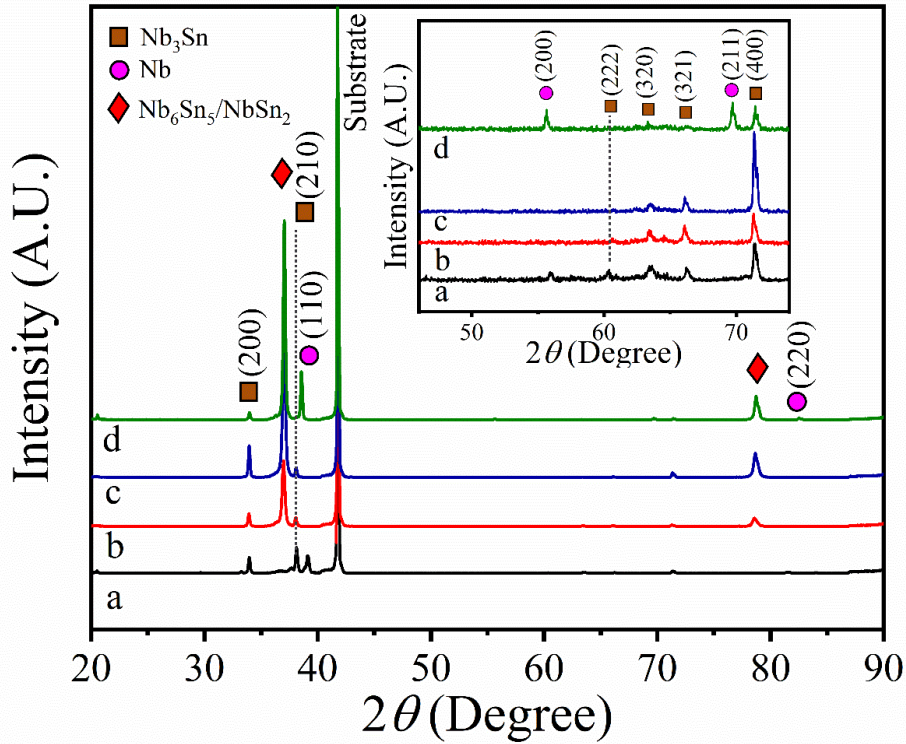

Fig. 1. XRD patterns of as-deposited and annealed films on sapphire substrate. Condition a, b, c, and d represent as-deposited film, and annealed films at 800, 900, and 1000 °C for 24 h, respectively. Inset shows the XRD pattern of the films at 45-75°.

The Raman spectra of the as-deposited films and films annealed for 24 h are shown in Fig. 2. The peaks were compared to previously reported Raman spectra of $Nb_3Sn$, Nb, $Nb_2O_5$, and $NbO_2$ [29-35]. The as-deposited film shows a wide Raman peak with its maximum at 189 cm$^{-1}$. This wide peak can be deconvolved into peaks at 134, 159, and 189 cm$^{-1}$. The peaks at 134 and 189 cm$^{-1}$ are close to the Raman lines assigned to the $F_{2g}$ and $E_g$ phonon modes of $Nb_3Sn$, respectively, as reported by Schicktanz et al. [30]. The $E_g$ optical phonon mode occurs in $A15$ compounds due to transition metal atoms that are lined up in chains, moving against each other along the chain. The wavelength of this mode was shown to shift with temperature [30]. In our results, the peak at 189 cm$^{-1}$ became stronger and shifted to 186 cm$^{-1}$ with annealing at 800 °C for 24 h. The peak shift from 189 to 186 cm$^{-1}$ is too small to make any conclusion about it, but we point out that the relief of residual stress in the film is known to cause Raman peak wavelength shift [36]. The $F_{2g}$ peak at 134 cm$^{-1}$ was suggested to be either due to an allowed mode in the $A15$ structure or due to defect-induced scattering [30]. The peak at 159 cm$^{-1}$ was previously observed in a near stoichiometric Nb$_3$Sn and was proposed to be possibly due to tetragonal micro-domains that are small in size to be observable in the obtained XRD patterns of Nb$_3$Sn, suggesting that Raman could be more sensitive than XRD in detecting micro-domain structures [30]. As the film is annealed to 900 °C, the peak at 186 cm$^{-1}$ is reduced in intensity, and an additional peak at ~216 nm appears. The peak at 216 cm$^{-1}$ is close to the reported peak position of NbO$_2$ [31] which could be resulted from the Sn evaporation from Nb$_3$Sn, potentially leaving only Nb on the surface, which may have oxidized as the sample got exposed to air. For films annealed at 1000 °C for 24 h, the Nb$_3$Sn peak at 186 cm$^{-1}$ (observed in Fig. 2(d) at 187 cm$^{-1}$) becomes small, due to decomposition of the Nb$_3$Sn and the of NbO$_2$ peak at 215 cm$^{-1}$ becomes the strongest peak. Two more peaks at 160 and 268 cm$^{-1}$ appeared as the film is annealed at 1000 °C for 24 h. These two peaks correspond to Nb$_3$Sn [30] and Nb [32] respectively. This result of the presence of Nb$_3$Sn and Nb after annealing at 1000 °C for 24 h agrees with the XRD data as both Nb and Nb$_3$Sn diffraction peaks are seen. These results, while preliminary, point out that Raman spectroscopy could be a powerful tool to study microstructural properties of Nb$_3$Sn films.

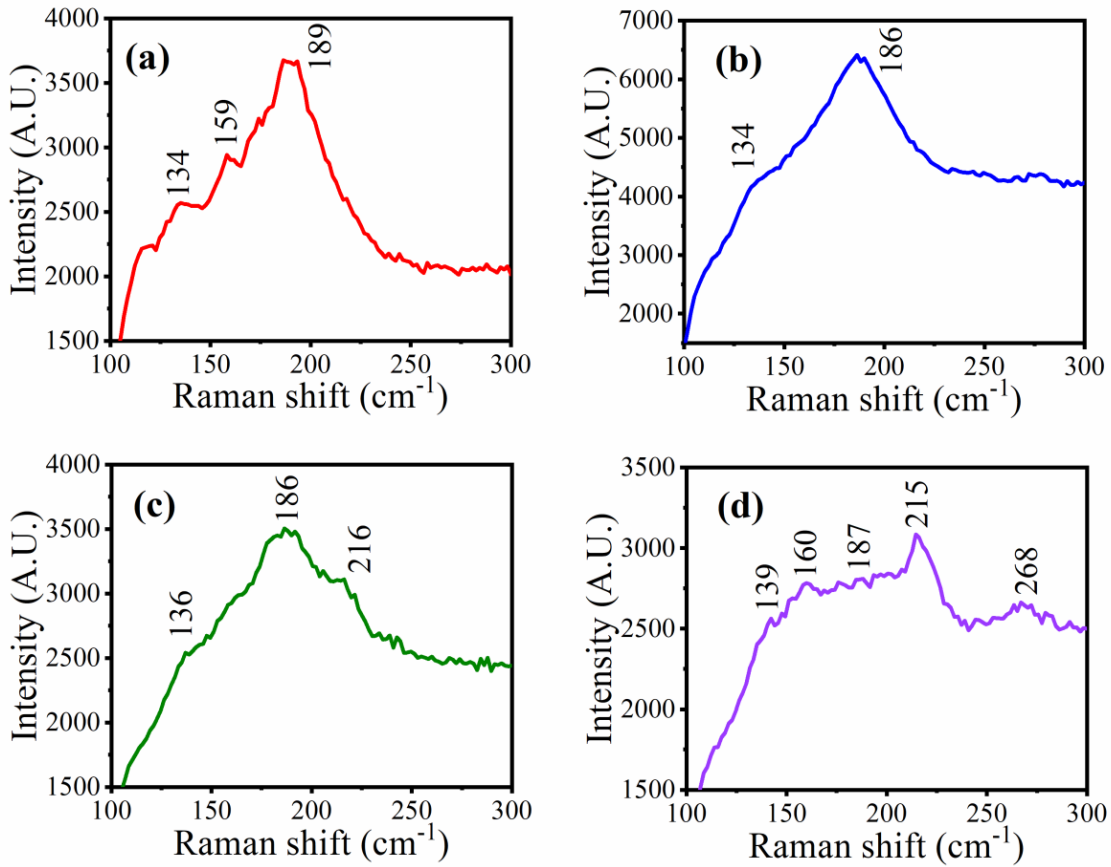

Fig. 2. Raman spectra of Nb$_3$Sn films on sapphire: (a) as-deposited, (b) annealed at 800 °C for 24 h, (c) annealed at 900 °C for 24 h, (d) annealed at 1000 °C for 24 h.

SEM micrographs of the surface of as-deposited and annealed films are shown in Fig. 3. The thickness of the as-deposited films after 6 h coating was ~350 nm, as estimated from the cross-sectional SEM image (inset of Fig. 3(a)). The surface of the as-deposited films consists of uniformly-distributed 50-150 nm grains with some randomly-distributed ~200-300 nm agglomerated clusters. The regular grains had ~27 at. % of Sn, whereas the clusters showed ~30 at. % of Sn, as detected by EDX. These values show that the clusters are Sn-rich, but the reported values are affected by the penetration of the EDX electron beam to deeper layers. As the films were annealed at 800, 900 and 1000 °C for 24 h, the grain size increased. Meanwhile, voids appeared on the surface (black regions in Fig. 3(b)). During the annealing, Sn evaporated from the surface. EDX analysis of the clusters showed ~19% Sn after annealing at 800 °C for 24 h, whereas the regular grain areas of the annealed film had ~21% Sn. The films had randomly-distributed particles on top of the grains after annealing at >800 °C. Fig. 3(e) and 3(f) show the surface of films annealed at 1000 °C for 1 and 12 h, respectively. The film annealed for 1 h had grains of ~100-150 nm diameter with some particles. As the annealing time was increased to 12 h, the grain size and the particle size on the films increased (Fig. 3(f)).

The film composition plays an important role in the superconducting properties of $Nb_3Sn$. Fig. 4 shows the Sn composition of ~350 nm thick films with annealing temperature and time, as measured by EDX. The as-deposited films had ~26.1 ± 0.5 at. % of Sn, which is consistent with the target composition of ~27 at. % measured prior to target placement in the sputter coater. For the annealing temperature of 1000 °C, Sn evaporates from the surface. We have observed large Sn loss when the film was annealed at 1000 °C for 24 h, with the Sn atomic composition reduced to ~4.1 ± 0.4 at. %. Sn loss during annealing was also observed on $Nb_3Sn$ films grown by multilayer sequential sputtering [16]. Sn loss was also observed on conventional vapor diffused $Nb_3Sn$ films after annealing at > 1100 °C [4]. The amount of Sn loss depends on both annealing temperature and annealing time. For the annealing temperature of 1000 °C, the Sn composition of the film was 11.3 ± 0.3 at. % for 12 h annealing time, and ~21.0 ± 0.6 at. % when the film was annealed for just 1 h.

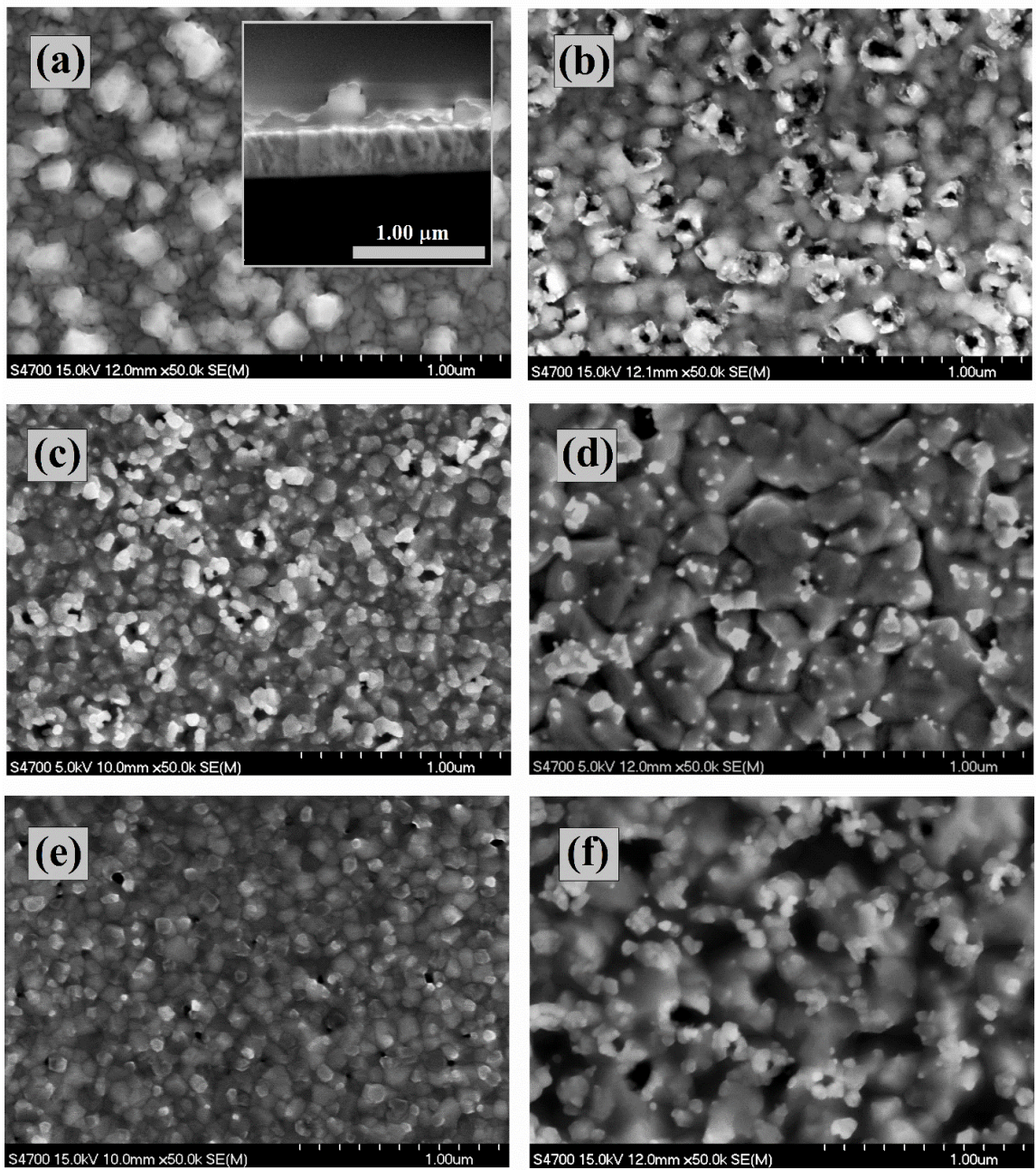

Fig. 3. SEM images of the surface of Nb$_3$Sn films deposited on sapphire: (a) as-deposited, (b) annealed at 800 °C for 24 h, (c) annealed at 900 °C for 24 h, (d) annealed at 1000 °C for 24 h, (e) annealed at 1000 °C for 1 h, and (f) annealed at 1000 °C for 12 h. The inset of (a) shows the cross-section of the as-deposited film.

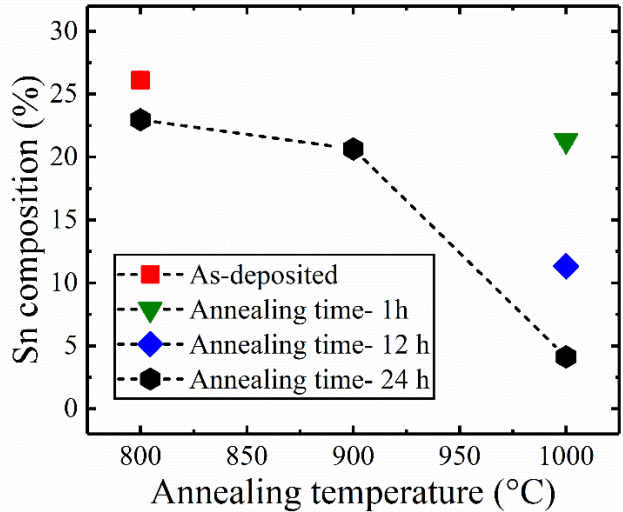

Fig. 4. Sn atomic composition of the Nb$_3$Sn films deposited on sapphire. The red square (■) is for the film deposited at 800 °C. The substrate heater was turned off right after deposition. The black hexagons (⬢) are for films deposited at 800 °C then annealed for 24 h. The blue diamonds (◆) and green triangles (▼) are for films annealed at 1000 °C for different times.

The AFM images of the films are shown in Fig. 5. The clusters observed in the SEM image of as-deposited films are more clearly observed in the AFM images. The lateral sizes of the clusters were ~300 nm, which is an agreement with those observed in the SEM images. The heights of the clusters were ~80-120 nm. The size and height of the regular grains increased after annealing whereas the height of the clusters did not change. As the films were annealed, the atoms gained enough activation energy to occupy the lower energy sites of the crystal lattice to form large grains [36, 37]. Fig. 6 shows RMS roughness obtained by averaging three different areas with error bars representing the standard deviation of measurements. The RMS roughness of the as-deposited films was ~20.2 ± 0.8 nm. The roughness of the films increased to ~29.7 ± 1.2 nm when the film was annealed at 800 °C for 24 h. Annealing at 900, and 1000 °C for 24 h increased roughness to 30.0 ± 0.4 and 32.2 ± 0.7 nm respectively. Films annealed at 1000 °C for 1, 12, and 24 h showed increased roughness with annealing time with most increase in roughness occurring after 12 h of annealing.

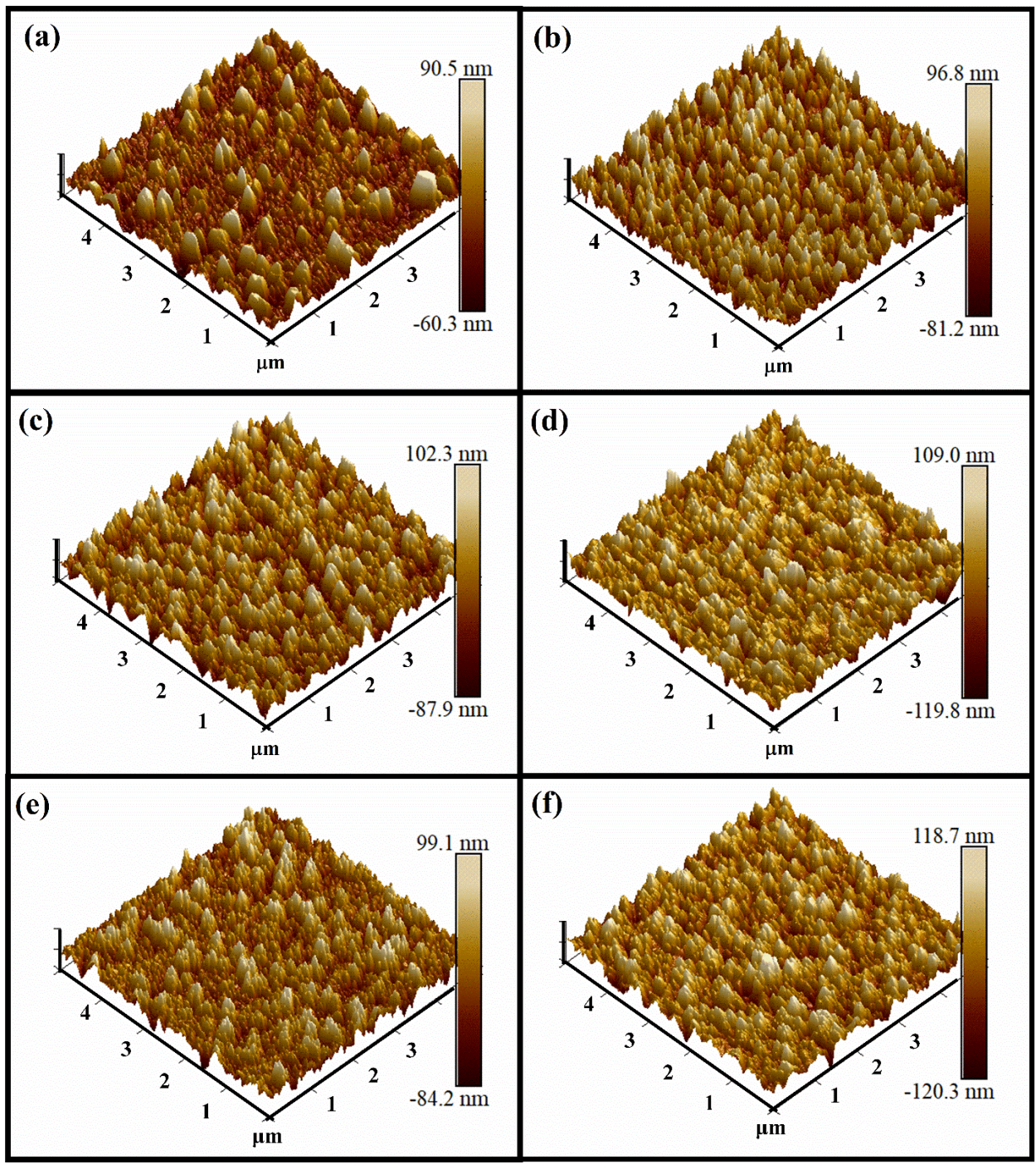

Fig. 5. AFM images of (5μm × 5 μm) scan size of Nb$_3$Sn films on sapphire: (a) as-deposited, (b) annealed at 800 °C for 24 h, (c) annealed at 900 °C for 24 h, (d) annealed at 1000 °C for 24 h, (e) annealed at 1000 °C for 1 h, and (f) annealed at 1000 °C for 12 h.

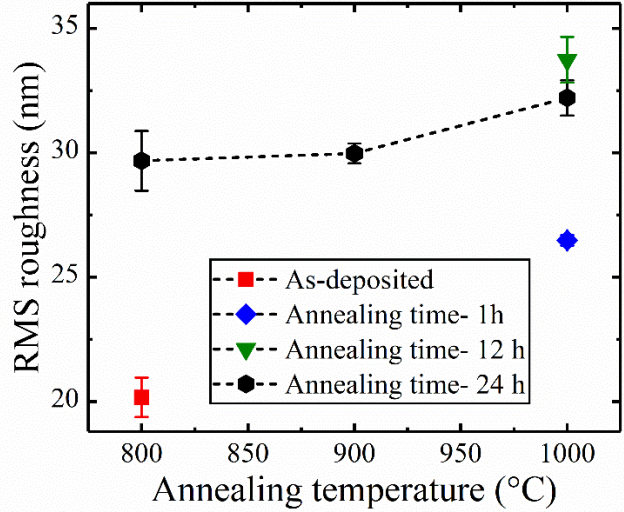

Fig. 6. RMS surface roughness of Nb$_3$Sn films on sapphire. The error bars represent the standard deviation of different measurements.

### 3.1.2. DC superconducting properties

Fig. 7(a) shows the film resistance versus temperature for as-deposited films and films annealed at 800, 900, and 1000 °C for 24 h, while Fig. 7(b) shows the resistance versus temperature for films annealed at 1000 °C for 1, 12, and 24 h. All films except the one annealed at 1000 °C for 24 h showed the superconducting transition of Nb$_3$Sn. The measured $T_c$, $\Delta T_c$, and *RRR* of the films are summarized in Table 1, along with the elemental Sn composition and surface roughness. The critical temperature $T_c$ of the as-deposited film was 17.21 K. $T_c$ increased with annealing at 800 °C for 24 h, reaching 17.83 K. Increasing the annealing temperature to 900 °C while keeping the annealing time constant for 24 h resulted in $T_c$ = 17.66 K. For films annealed at 1000 °C, the Sn composition decreased with increased annealing time, consequently degrading superconducting properties. For the film annealed at 1000 °C for 24 h, the surface resistance starts to drop near 17.7 K (position A in the inset of Fig. 7(a)) and a sharp drop of resistance is observed near 9 K (position B in the inset of Fig. 7(a)). This behavior is due to the superconducting transitions of Nb ($T_c$ ~9.3 K) and Nb$_3$Sn ($T_c$ ~18 K) present in the film. The transition width increased with increased annealing temperature and time. Based on the data, high quality superconducting films can be obtained with annealing at 800-900 °C for 24 h or annealing at 1000 °C for 1 h.

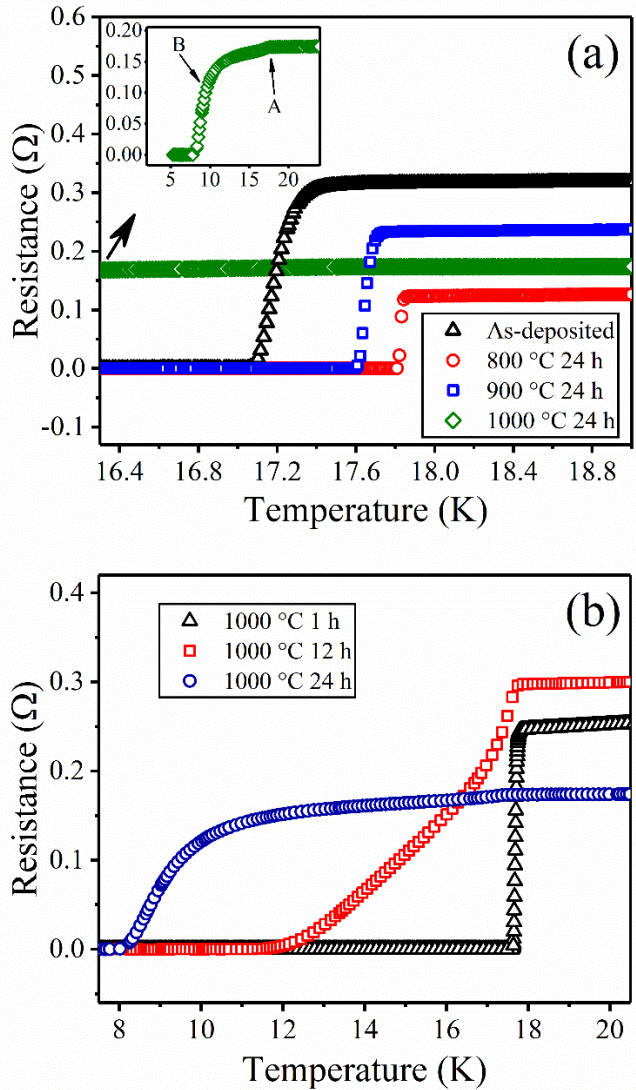

Fig. 7. (a) Resistance versus temperature of as-deposited film and films annealed at three different temperature for 24 h, the inset of shows the transition region of film annealed at 1000 °C for 24 h, (b) resistance versus temperature of films annealed at 1000 °C for 1, 12, and 24 h.

The residual resistivity ratio *RRR* calculated as a ratio of the surface resistance at 300 and at 20 K is shown in Table 1. The as-deposited film had a *RRR* value of 2.53. The highest *RRR* value of 5.41 was observed on the film annealed at 800 °C for 24 h. Further increase in annealing temperature reduced the *RRR* of the film. For the annealing temperature of 1000 °C, the *RRR* was reduced from 3.01 to 2.07 when the annealing time was increased from 1 to 24 h.

Table 1. Film properties of as-deposited and annealed films on sapphire substrate. The films were coated at 800 °C coating temeprature for 6 h.

| Annealing temperature (°C) | Annealing time (h) | Sn composition (%) | RMS roughness (nm) | $T_c$ (K) | $\Delta T_c$ (K) | $RRR$ |
|---|---|---|---|---|---|---|
| NA | NA | 26.1 ± 0.5 | 20.2 ± 0.8 | 17.21 | 0.18 | 2.53 |
| 800 | 24 | 23.0 ± 0.3 | 29.7 ± 1.2 | 17.83 | 0.03 | 5.41 |
| 900 | 24 | 20.6 ± 0.4 | 30.0 ± 0.4 | 17.66 | 0.06 | 4.36 |
| 1000 | 1 | 21.0 ± 0.6 | 26.5 ± 0.2 | 17.68 | 0.06 | 3.01 |
| 1000 | 12 | 11.3 ± 0.3 | 33.7 ± 0.9 | 15.74 | 3.61 | 2.13 |
| 1000 | 24 | 4.1 ± 0.4 | 32.2 ± 0.7 | 10.95 | 4.69 | 2.07 |

## 3.2. Nb$_3$Sn deposited on Nb

### 3.2.1. Film structure, morphology, and composition

The X-ray diffraction patterns of the Nb$_3$Sn films grown on Nb substrates are shown in Fig. 8. The XRD peaks due to the Nb substrate are shown in Fig. 8(a). All films showed diffraction patterns for Nb$_3$Sn with no diffraction peaks corresponding to Nb$_6$Sn$_5$ or NbSn$_2$ detected. The peak intensity of the (211) diffraction order increased after annealing. The annealed films showed an additional diffraction peak near the (210) reflection. The $2\theta$ value for this peak is that of the Nb (110) diffraction order. The peak intensity of this diffraction order increased upon annealing at 1000 °C and was accompanied by a reduction in the Nb$_3$Sn (210) diffraction order peak intensity. The crystallite size estimated from the (200), (210), and (211) diffraction orders based on the Scherrer equation [38] are shown in Fig. 8(e). The crystallite size increases in a narrow range from 32 to 41 nm after annealing.

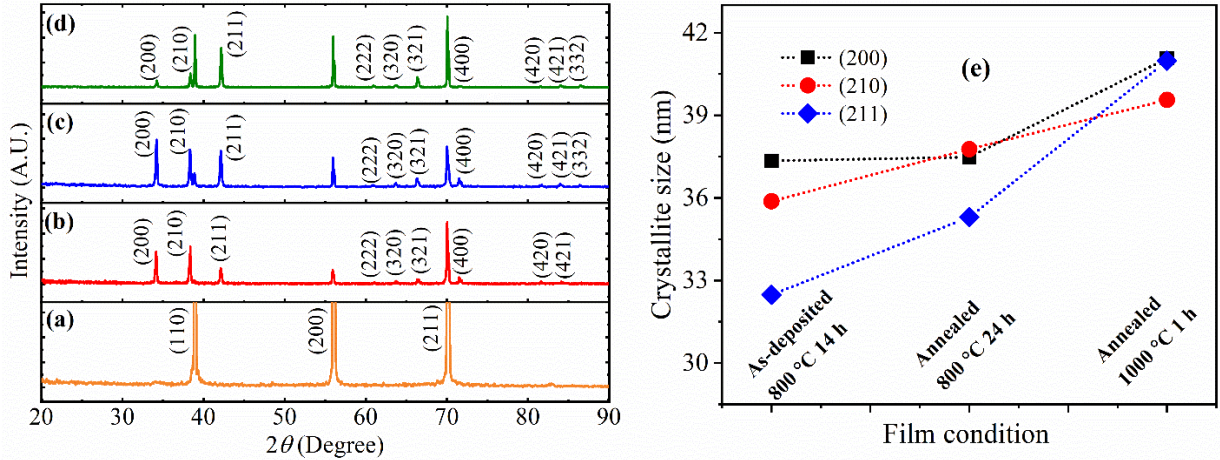

Fig. 8. (a-d): XRD patterns of (a) Nb substrate, (b) as-deposited film, (c) film annealed at 800 °C for 24 h, and (d) film annealed at 1000 °C for 1 h, (e) crystallite size of different crystal planes of the as-deposited and annealed films.

Fig. 9 shows the Raman spectra of the as-deposited $Nb_3Sn$ film (a) and films annealed at 800 °C for 24 h (b) and at 1000 °C for 1 h (c). The Raman spectra of the as-deposited film had a wide peak that can be deconvoluted into $F_{2g}$ and $E_g$ phonon modes of $Nb_3Sn$ at 138 and 192 cm$^{-1}$. The peaks shifted slightly after annealing. The peak at 192 cm$^{-1}$ shifted to 189 cm$^{-1}$ when annealed at 800 °C for 24 h and to 186 cm$^{-1}$ when annealed at 1000 °C for 1 h, whereas the peak at 138 cm$^{-1}$ shifted to 136 and 134 cm$^{-1}$ for 800 °C and 1000 °C, respectively. A notable difference between the as-deposited films grow on sapphire and Nb is that the film on Nb (Fig. 9(a)) does not show a peak at 159 cm$^{-1}$ as observed for the film grown on sapphire (Fig. 2(a)), which could be interpreted that tetragonal micro-domains of $Nb_3Sn$ are not formed on the Nb substrate. Also, for the annealing conditions in Fig, 9, no Raman peaks for Nb or $NbO_2$ were observed.

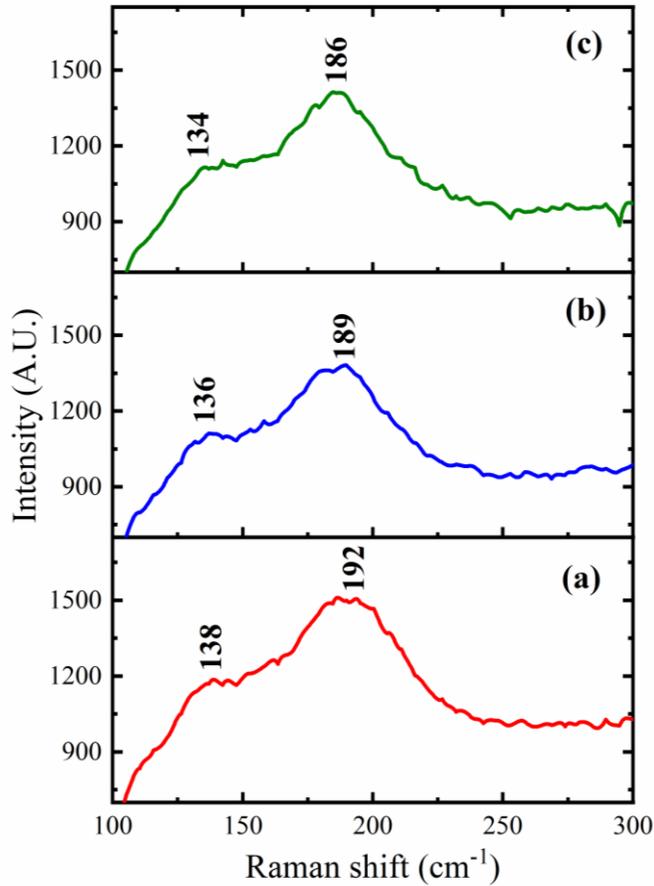

Fig. 9. Raman spectra of (a) as-deposited films, (b) film annealed at 800 °C for 24 h, and (d) film annealed at 1000 °C for 1 h.

Fig. 10 shows the optical and SEM images of $Nb_3Sn$ films on Nb. Similar to the $Nb_3Sn$ films deposited on sapphire, the films on Nb substrates also have scattered clusters on the surface. The distributions of these clusters seem to depend on the grain orientation of the Nb substrate, as seen in Fig 10(a) and 10(d). The clustered density was reduced after annealing as observed in the optical and SEM images in Fig. 10(a-f). The grain size and microstructure of the as-deposited film differed from those of the annealed films. As seen in Fig. 10(g-i), as-deposited films had faceted small grains, whereas the annealed films showed rounded grains with increased grain size.

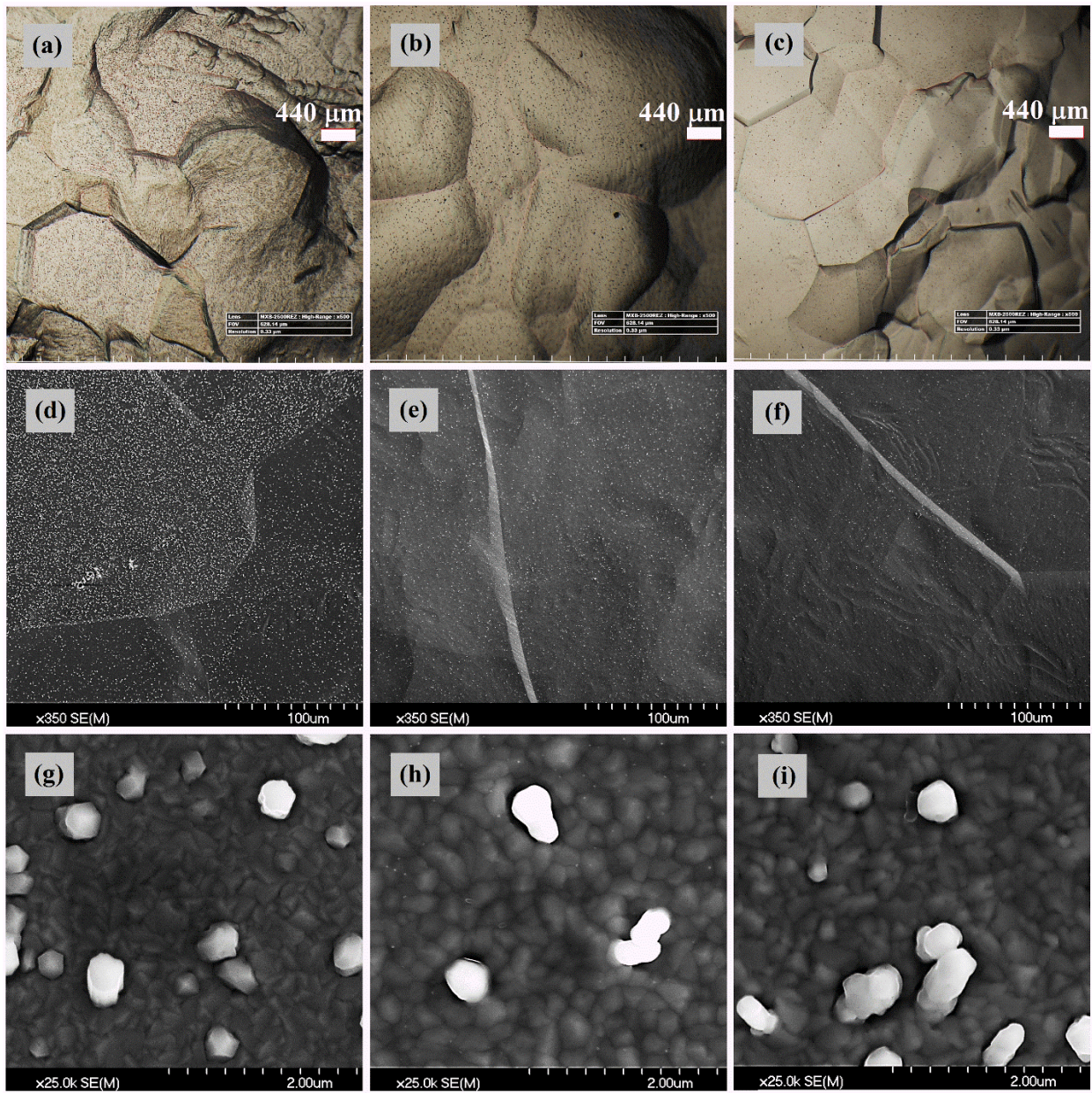

Fig. 10. (a-c): Surface micrographs under an optical microscope (a) as-deposited film, (b) film annealed at 800 °C for 24 h, (c) film annealed at 1000 °C for 1 h, (d-f): SEM micrograph of the surface with ×350 magnification (d) as-deposited film, (e) film annealed at 800 °C for 24 h, (f) film annealed at 1000 °C for 1 h, (g-i): SEM micrograph of the surface with ×25000 magnification (g) as-deposited film, (h) film annealed at 800 °C for 24 h, (i) film annealed at 1000 °C for 1 h.

To estimate the Sn composition on the regular and clustered grains, EDX spot analysis was performed. Fig. 11 shows the spot analysis of the as-deposited film. Both regular and clustered grains had similar Sn composition of ~ 23-24.5 at. % which is different than those Sn-rich clusters observed on sapphire substrates. For films annealed at 800 °C for 24 h, the Sn concentration on the clustered grain was ~23.6 %, whereas for the regular grain area was ~22.7 at. %. For the films annealed at 1000 °C for 1 h, Sn composition on clustered and regular grains were ~21.2 at. % and ~21.6 at. % respectively. The overall Sn atomic % on the 1.2 mm$^2$ surface area were ~22.3 ± 0.3 % for the as-deposited film, 22.9 ± 0.7 % for the film annealed at 800 °C for 24 h, and 22.4 ± 0.2 % for the film annealed at 1000 °C for 1 h. The contrast difference in SEM images, between regular grains and clusters may have come from the topographic variations with the clusters having similar composition as the grains but with higher heights.

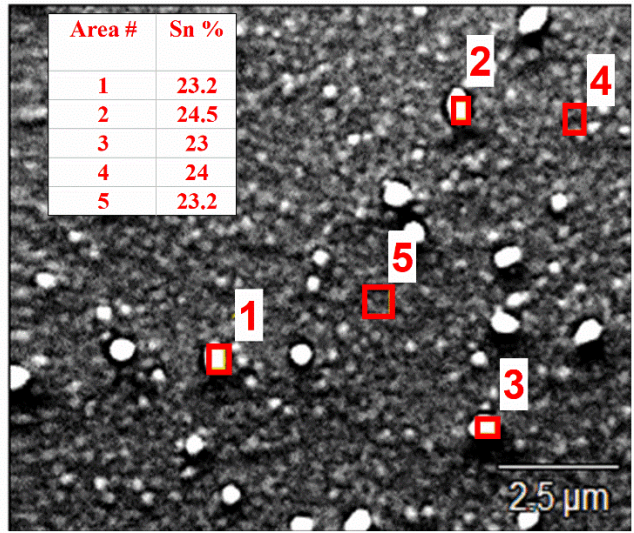

Fig. 11. EDX spot analysis of five different regions of as-deposited Nb$_3$Sn film surface. The Sn atomic % of the spots are given in the inset.

Fig. 12 shows the 2D and 3D AFM images and line scans taken along the lines shown on the 2D images of the films. The random distribution of the clustered grains of different heights gave rise to film surface roughness. The height of these irregular clustered grains in as-deposited films was in the range of ~50-300 nm whereas the regular grains of the films had height below 40 nm as seen in Fig. 12(a). The average height of 10 different clusters of the film annealed at 800 °C for 24 h was 132 ± 41 nm, whereas the film annealed at 1000 °C for 1 h had an average height of 190 ± 88 nm.

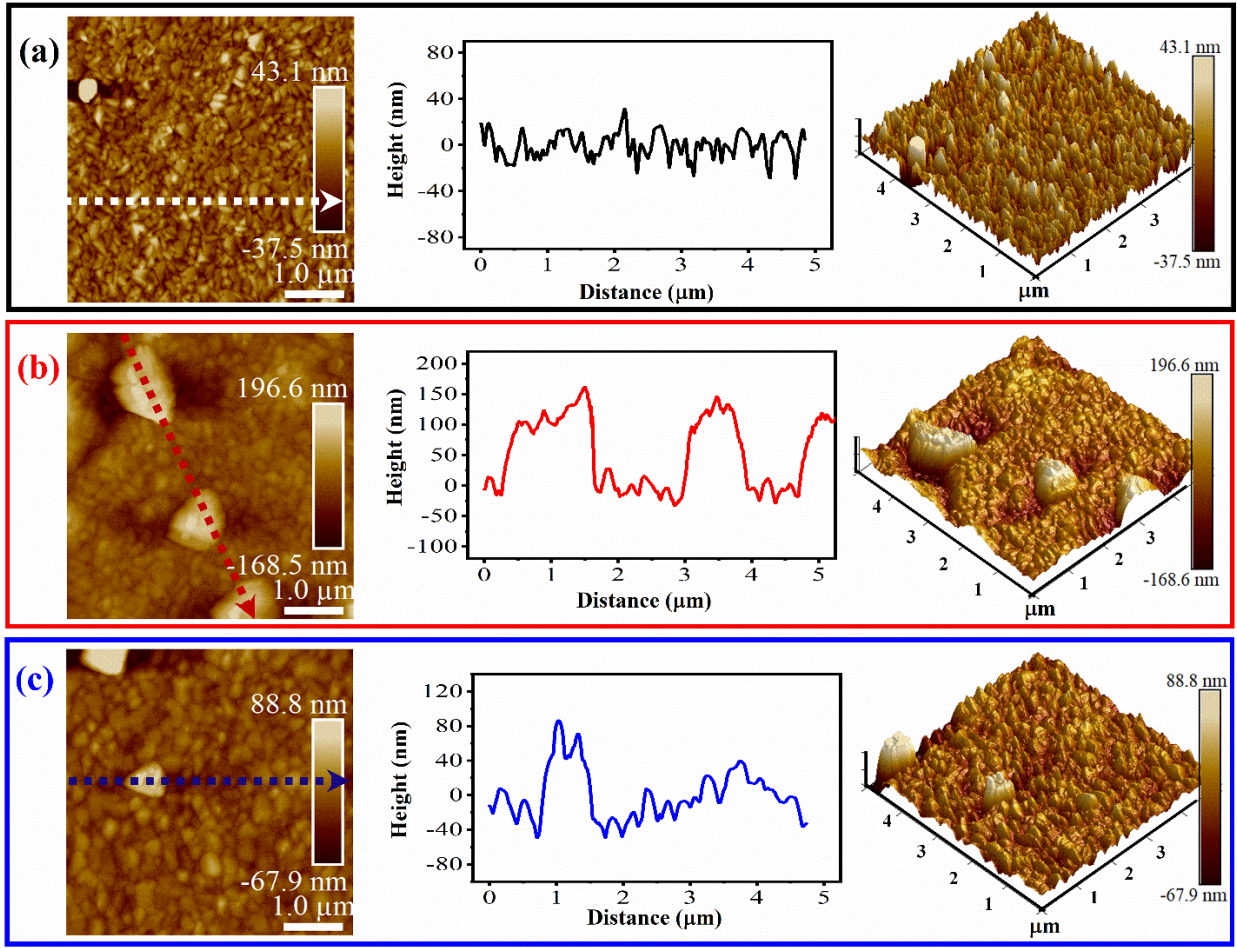

Fig. 12. 2D and 3D AFM images of Nb$_3$Sn films on Nb substrate: (a) as-deposited, (b) annealed at 800 °C for 24 h, and (c) annealed at 1000 °C for 1 h. The line analysis of the surface corresponds to the broken arrows on 2D images are shown in the middle.

### 3.2.2. RF Superconducting Properties

Fig. 13 shows the RF surface resistance of the as-deposited film, and film annealed at 800 °C for 24 h and at 1000 °C for 1 h. The error bars were estimated from fluctuations in ambient temperature and pressure, noise in the electronics, and calibration errors during the measurement [39]. The inset of Fig. 13 shows the surface resistance for all three conditions as a function of 1/temperature over the range 6-23 K. The surface resistance $R_{RF}$ is obtained from the power dissipation induced by an applied RF field [40]:

$$R_{RF}(H_P, T_S) = \frac{P_{RF}(H_P, T_S)}{k \cdot H_P^2} \qquad (1)$$

Here, $R_{RF}$, $P_{RF}$, $H_P$, and $T_S$ are RF surface resistance, power dissipation induced by the RF field, peak magnetic field on the surface of the sample, and sample temperature, respectively. The geometrically-dependent constant $k = 3.7 \times 10^7$ W/$\Omega$T$^2$. The RF induced power was calculated from the power compensation technique where $P_{RF}$ is the difference between the heater power required to keep the sample temperature unchanged without the RF field in the cavity and the heater power required to keep the sample temperature unchanged with the RF field [40]:

$$P_{RF}(H_P, T_S) = P_h^{H=0, T=T_s} - P_h^{H=H_P, T=T_s} \qquad (2)$$

Here, $P_h^{H=0, T=T_s}$ is the heater power required to keep the sample temperature $T_S$ unchanged without the RF field, and $P_h^{H=H_P, T=T_s}$ is the heater power required to keep the sample temperature $T_S$ unchanged with the RF field $H_P$.

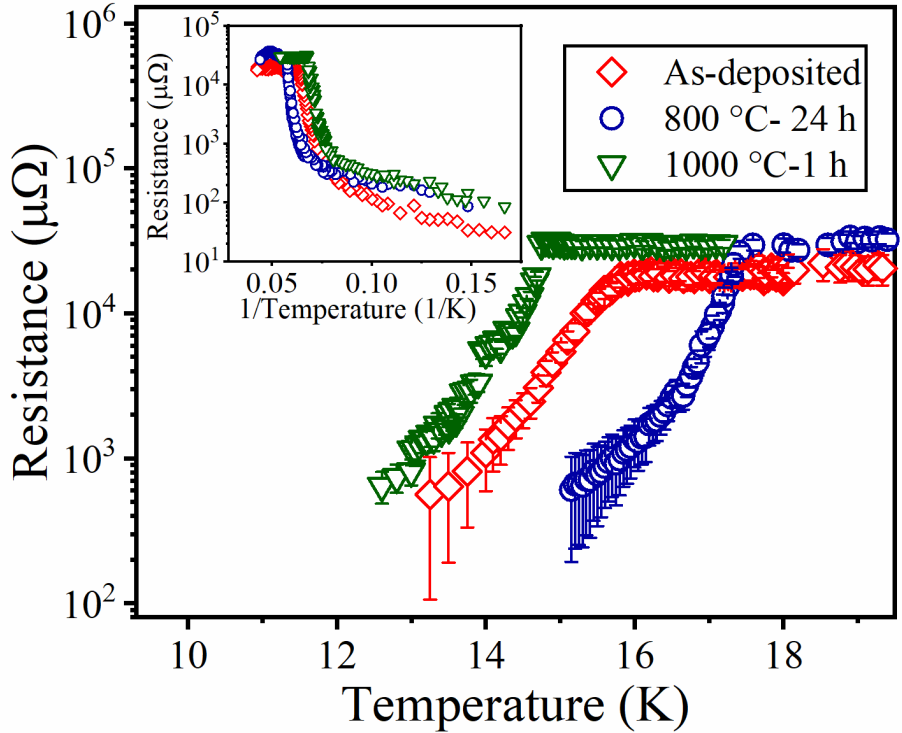

Fig. 13. RF surface resistance of the Nb$_3$Sn films near critical temperature. The inset shows the resistance vs 1/temperature over 6-23 K temperature range.

From Fig. 13, both as-deposited and annealed films had a superconducting transition due to the Nb$_3$Sn films. The resistance of as-deposited films dropped sharply below 16.02 K. The $T_c$ obtained from the as-deposited film was 16.02 K, which increased to 17.44 K after annealing at 800 °C for 24 h. The RF superconducting properties of the film degraded when annealed at 1000 °C for 1 h. For this condition, $T_c$ was 14.22 K. For all samples, no transition was observed near 9.3 K which confirmed that the RF field did not penetrate to the Nb substrate. The RF surface resistance at ~18.5 K in the normal conducting state increased from 22.2 ± 5.5 to 29.7 ± 3 mΩ when the as-deposited film was annealed at 800 °C for 24 h. The measured surface resistance of the film annealed at 1000 °C for 1 h was 29.9 ± 2 mΩ at ~18.5 K.

## 4. Conclusion

We fabricated $Nb_3Sn$ films on sapphire and Nb substrates by DC magnetron sputtering from a stoichiometric target. The films were post-annealed at different temperatures for different durations. The as-deposited film had polycrystalline $Nb_3Sn$ and the crystallinity of the film improved when annealed at 800 °C for 24 h. The film quality degraded when annealed at 1000 °C for 12 and 24 h due to Sn evaporation from the surface. The highest $T_c$ =17.83 K was obtained from the film annealed at 800 °C for 24 h. The RF properties of films deposited on Nb substrates were also measured. Both as-deposited and annealed films had superconducting transition due to $Nb_3Sn$. The $T_c$ of the as-deposited film increased from 16.02 K to 17.44 K when annealed at 800 °C for 24 h, and decreased to 14.22 K when annealed at 1000 °C for 1 h. Based on our RF results, magnetron sputtering from a stoichiometric $Nb_3Sn$ target can be used to coat large cavities and the performance of the coated cavities can be improved by heat treatment with best conditions from the those investigated is annealing at 800 °C for 24 h. Annealing at 1000 °C reduced Sn content due to Sn evaporation from the $Nb_3Sn$ layer adversely affecting the superconducting properties of the film. Raman study of the films showed high sensitivity to the surface composition and showed that, the strength of the $Nb_3Sn$ $E_g$ phonon modes and absence of $NbO_2$ and Nb modes on the film annealed at 800 °C for 24 correlates well with obtaining the highest $T_c$ for the film.

**Acknowledgement**

This material is based upon work supported by the U.S. Department of Energy, Office of Science, Office of Nuclear Physics under U.S. DOE Contract No. DE-AC05-06OR23177. Some of the characterizations are performed at the Applied Research Center Core Labs at The College of William and Mary. The authors would like to thank Dr. Gianluigi Ciovati for his comments. The authors acknowledge Peter Owen and Pete Kushnick of Jefferson Lab for their continuous help on RF surface resistance measurement and Joshua Spradlin of Jefferson Lab for his help with $T_c$ measurement.

**Reference**

[1] P. Seidel ed., Applied Superconductivity: Handbook on Devices and Applications Wiley-VCH Verlag GmbH & Co. KGaA, (2015). http://doi.org/10.1002/9783527670635.

[2] U. Pudasaini, G. Eremeev, C. E. Reece, J. Tuggle, and M. Kelley, Initial growth of tin on niobium for vapor diffusion coating of $Nb_3Sn$, *Supercond. Sci. Technol.* 32 (2019) 045008. https://doi.org/10.1088/1361-6668/aafa88.

[3] S. Posen, D.L. Hall, $Nb_3Sn$ superconducting radiofrequency cavities: fabrication, results, properties, and prospects, *Supercond. Sci. Technol.* 30 (2017) 033004. https://doi.org/10.1088/1361-6668/30/3/033004.

[4] U. Pudasaini, G. Ciovati, G. V. Eremeev, M. J. Kelley, I. Parajuli, C. E. Reece, M. N. Sayeed, Recent Results From Nb$_3$Sn Single Cell Cavities Coated at Jefferson Lab. in *Proc. SRF'19* (2019) 65-70. http://jacow.org/srf2019/papers/mop018.pdf.

[5] U. Pudasaini, M. J. Kelley, G. Ciovati, G. V. Eremeev, C. E. Reece, I. P. Parajuli, M. N. Sayeed, Nb$_3$Sn multicell cavity coating at JLAB, in *Proc. IPAC'18,* Vancouver, Canada, (2018) 1798-1803. https://doi.org//10.18429/JACoW-IPAC2018-WEYGBF3.

[6] G. Eremeev, W. Clemens, K. Macha, C.E. Reece, A.M. Valente-Feliciano, S. Williams, U. Pudasaini, and M. Kelley, Nb$_3$Sn multicell cavity coating system at JLAB. (2020) arXiv preprint arXiv:2001.03823.

[7] S. Posen et al., Nb$_3$Sn at Fermilab: Exploring Performance, in *Proc. SRF'19*, Dresden, Germany, (2019), 818-822. https://doi.org/10.18429/JACoW-SRF2019-THFUB1

[8] R.D. Porter, M. Liepe, and J.T. Maniscalco, "High Frequency Nb$_3$Sn Cavities", in *Proc. SRF'19*, Dresden, Germany, (2019) 44-47. https://doi.org10.18429/JACoW-SRF2019-MOP011.

[9] M. Perpeet, A. Cassinese, M. A. Hein, T Kaiser, G. Muller, H. Piel, J. Pouryamout, Nb$_3$Sn films on sapphire. A promising alternative for superconductive microwave technology.. 9(2) (1999) 2496-2499, https://doi.org/10.1109/77.784987.

[10] C.T. Wu, R.T. Kampwirth, J.W. Hafstrom, High-rate magnetron sputtering of high $T_c$ Nb$_3$Sn films, *J. Vac. Sci. Technol.* 14 (1977) 134-137. https://doi.org/10.1116/1.569104.

[11] R. Kampwirth, J. Hafstrom, C. Wu, Application of high rate magnetron sputtering to the fabrication of A-15 compounds, *IEEE Transactions on Magnetics*, 13 (1977) 315-318. https://doi.org/10.1109/TMAG.1977.1059445.

[12] K. Agatsuma, H. Tateishi, K. Arai, T. Saitoh, M. Nakagawa, Nb$_3$Sn thin films made by RF magnetron sputtering process with a reacted Nb$_3$Sn powder target, *IEEE Transactions on Magnetics* 32 (1996) 2925-2928. https://doi.org/10.1109/20.511488.

[13] E.A. Ilyina, G. Rosaz, J.B. Descarrega, W. Vollenberg, A.J.G. Lunt, F. Leaux, S. Calatroni, W. Venturini-Delsolaro, M. Taborelli, Development of sputtered Nb$_3$Sn films on copper substrates for superconducting radiofrequency applications, *Supercond. Sci. Technol.* 32 (2019) 035002. https://doi.org/10.1088/1361-6668/aaf61f.

[14] J. Vandenberg, M. Gurvitch, R. Hamm, M. Hong, J. Rowell, New phase formation and superconductivity in reactively diffused Nb$_3$Sn multilayer films, *IEEE transactions on magnetics*, 21 (1985) 819-822. https://doi.org/10.1109/TMAG.1985.1063612.

[15] A.A. Rossi, S.M. Deambrosis, S. Stark, V. Rampazzo, V. Rupp, R.G. Sharma, F. Stivanello, V. Palmieri, Nb$_3$Sn films by multilayer sputtering, in *Proc. SRF'09* Berlin, Germany, (2009) 149-154. http://accelconf.web.cern.ch/AccelConf/SRF2009/papers/tuobau06.pdf.

[16] M. N. Sayeed, U Pudasaini, C. E. Reece, G. Eremeev, H. E. Elsayed-Ali. Structural and superconducting properties of Nb$_3$Sn films grown by multilayer sequential magnetron sputtering, *J. Alloys Compd.* 800 (2019) 272-278. https://doi.org/10.1016/j.jallcom.2019.06.017.

[17] C. Sundahl Synthesis of Superconducting Nb$_3$Sn Thin Film Heterostructures for the Study of High-Energy RF Physics, PhD Dissertation, (2019), University of Wisconsin-Madison.

[18] R. Valizadeh et al., PVD Depostion of Nb$_3$Sn Thin Film on Copper Substrate from an Alloy Nb$_3$Sn Target, in *Proc. IPAC'19*, Melbourne, Australia, (2019), 2818-2821. https://doi.org/10.18429/JACoW-IPAC2019-WEPRB011.

[19] L. Xiao, X.Y. Lu, W.W. Tan, D. Xie, Y. Yang, and L. Zhu, The Technical Study of Nb$_3$Sn Film Deposition on Copper by HiPIMS, in *Proc. SRF'19*, Dresden, Germany, (2019) 846-847, https://doi.org/10.18429/JACoW-SRF2019-THP008.

[20] M. N. Sayeed et al., Deposition of Nb$_3$Sn Films by Multilayer Sequential Sputtering for SRF Cavity Application, in *Proc. SRF'19*, Dresden, Germany, (2019) 637-641. https://doi.org/10.18429/JACoW-SRF2019-TUP079.

[21] J. Li, Magnetron sputtering of Superconducting Multilayer Nb$_3$Sn Thin Film (Master thesis), (2009), University of Padua and National Institute for Nuclear Physics (INFN), Legnaro, Italy.

[22] G. Eremeev, B. Clemens, K. Macha, H. Park, R.S. Williams, Development of a Nb$_3$Sn cavity vapor diffusion deposition system, In *Proc. SRF'13* Paris, France, (2013) 603-606. http://accelconf.web.cern.ch/AccelConf/SRF2013/papers/tup071.pdf.

[23] M.N. Sayeed, U. Pudasaini, C.E. Reece, G. V. Eremeev, H.E. Elsayed-Ali, Effect of layer thickness on structural, morphological and superconducting properties of Nb$_3$Sn films fabricated by multilayer sequential sputtering, *IOP Conf. Ser.: Mater. Sci. Eng.* https://doi.org/10.1088/1757-899X/756/1/012014.

[24] L. Allen, M. Beasley, R. Hammond, J. Turneaure, RF surface resistance of Nb$_3$Sn, NbZr, and NbN thin films. *IEEE transactions on magnetics*. 23(2) (1987) 1405-1408. https://doi.org/10.1109/TMAG.1987.1064870.

[25] A. Andreone, A. Cassinese, A. Di Chiara, M. Iavarone, F. Palomba, A. Ruosi, R. Vaglio, Nonlinear microwave properties of Nb$_3$Sn sputtered superconducting films, *J. Appl. Phys*. 82 (1997) 1736-1742. https://doi.org/10.1063/1.365975.

[26] M. Arzeo et al., RF Performances of Nb$_3$Sn Coatings on a Copper Substrate for Accelerating Cavities Applications, presented at the *19th Int. Conf. on RF Superconductivity (SRF'19)*, Dresden, Germany, Jun.-Jul. 2019, paper THFUA7.

[27] J. Spradlin, A.-M. Valente-Feliciano, C. Reece, A multi-sample residual resistivity ratio system for high quality superconductor measurements, in *Proc. SRF'15,* Whistler, Canada (2015) 726-730. https://doi.org/10.18429/JACoW-SRF2015-TUPB063.

[28] B. P. Xiao, C.E. Reece, H.L. Phillips, R. L. Geng, H. Wang, F. Marhauser, M.J. Kelley, Note: Radio frequency surface impedance characterization system for superconducting samples at 7.5 GHz, *Rev. Sci. Inst.*, 82 (5) (2011) 056104. https://doi.org/10.1063/1.3575589.

[29] S.B. Dierker, M.V. Klein, G.W. Webb, and Z. Fisk, Electronic Raman scattering by superconducting-gap excitations in Nb$_3$Sn and V$_3$Si. *Physical Review Letters*, 50(11) (1983) 853. https://doi.org/10.1103/PhysRevLett.50.853.

[30] S. Schicktanz, R. Kaiser, E. Schneider, and W. Gläser, Raman studies of A15 compounds. Physical Review B, 22(5) (1980) 2386. https://doi.org/10.1103/PhysRevB.22.2386.

[31] S.H. Lee, H.N. Yoon, I.S. Yoon, and B.S. Kim, Single crystalline NbO$_2$ nanowire synthesis by chemical vapor transport method. Bulletin of the Korean Chemical Society, 33(3) (2012) 839-842. http://dx.doi.org/10.5012/bkcs.2012.33.3.839.

[32] N. Singh, M.N. Deo, M. Nand, S.N. Jha, and S.B. Roy, Raman and photoelectron spectroscopic investigation of high-purity niobium materials: Oxides, hydrides, and hydrocarbons. Journal of Applied Physics, 120(11) (2016) 114902. https://doi.org/10.1063/1.4962650.

[33] A.M. Raba, J. Bautista-Ruíz, and M.R. Joya, Synthesis and structural properties of niobium pentoxide powders: a comparative study of the growth process. Materials Research, 19(6) (2016) 1381-1387. http://dx.doi.org/10.1590/1980-5373-mr-2015-0733.

[34] J. Zasadzinski, B. Albee, S. Bishnoi, C. Cao, G. Ciovati, L.D. Cooley, D.C. Ford, and T. Proslier, 2011. Raman Spectroscopy as a Probe of Surface Oxides and Hydrides on Niobium, in *Proc. SRF'11* Chicago, USA (2011) 912-916.

[35] I.A. Alhomoudi, G. Newaz, Residual stresses and Raman shift relation in anatase TiO$_2$ thin film, *Thin Solid Films*, 517(15) (2009) 4372-4378.

[36] Y. Chen, N. Jyoti, K. Hyun-U, J. Kim. Effect of annealing temperature on the characteristics of ZnO thin films. *Journal of Physics and Chemistry of Solids*, 73(11) (2012) 1259-1263. https://doi.org/10.1016/j.jpcs.2012.06.007.

[37] J. Sengupta, R.K. Sahoo, K.K. Bardhan, C.D. Mukherjee, Influence of annealing temperature on the structural, topographical and optical properties of sol–gel derived ZnO thin films, *Mat. Let.* 65 (2011) 2572-2574. https://doi.org/10.1016/j.matlet.2011.06.021.

[38] B. Cullity, Elements of X-Ray Diffraction 2nd edition. Addision-Wesley Pub. Co, Inc., CA, USA, 197, (1978) 356.

[39] G. Eremeev, Preliminary error analysis for surface impedance characterization (SIC) system, *Technical Report JLAB-TN-12-048*, TJNAF (2012).

[40] G. V. Eremeev, H. L. Phillips, A-M. Valente-Feliciano, C. E. Reece, and B. P. Xiao, Characterization of Superconducting Samples With SIC System for Thin Film Developments: Status and Recent Results, in *Proc. SRF'13* Paris, France, (2013) 599-602.